\documentclass{article}
\pdfoutput=1 
\usepackage{times}
\usepackage{geometry}
\usepackage{titlesec}
\usepackage{setspace} 
\usepackage{ragged2e}
\usepackage{hyperref}
\usepackage{graphicx}
\usepackage{caption}
\usepackage{hhline}
\usepackage{fancyhdr}
\usepackage{ifthen}
\usepackage{url}
\usepackage{footmisc}
\usepackage{xcolor}
\usepackage{etoolbox}
\usepackage{textcase}
\usepackage{tabularx} 
\usepackage{amsmath}
\usepackage{wrapfig}
\usepackage{subcaption}

\definecolor{lightgraytext}{gray}{0.4} 

\DeclareCaptionLabelFormat{figlabel}{{FIG. #2. }}
\DeclareCaptionLabelFormat{tablelabel}{{TABLE #2. }}
\titleformat{\section}
  {\fontsize{10}{12}\selectfont\MakeUppercase} 
  {\thesection.} 
  {0.5em} 
  {} 

\titleformat{\subsection}
  {\fontsize{10}{12}\selectfont\bfseries} 
  {\thesubsection.} 
  {0.5em} 
  {} 

\titleformat{\subsubsection}
  {\fontsize{10}{12}\selectfont\itshape} 
  {\thesubsubsection.} 
  {0.5em} 
  {} 

\newcommand{\centeredparheader}[1]{%
    \textcolor{lightgraytext}{%
        \parbox[c]{\textwidth}{%
            \centering #1%
        }%
    }%
} 

\fancypagestyle{plain}{
  \fancyhf{}%
  \fancyhead[C]{\centeredparheader{%
    \fontsize{8}{0}{\textbf{DATTA et al.}} \quad\textcolor{lightgraytext}{\\ 
    }}}%
}

\newcommand{\evenheader}{\centeredparheader{%
   \fontsize{8}{0}{\textbf{IAC/P9-3}} \quad\textcolor{lightgraytext}{\\ 
   }}}
    
\newcommand{\oddheader}{\centeredparheader{%
    \fontsize{8}{0}{\textbf{DATTA et al.}} \quad\textcolor{lightgraytext}{\\ 
    }}}
    
\begin{document}

\title{\raggedright
  \textbf{\fontsize{12}{15}\selectfont
  \MakeUppercase{WHOLE DEVICE MODELING OF THE FUZE SHEARED-FLOW-STABILIZED Z \\
  PINCH 
}}
  \vspace{-3.2em}
  \setstretch{0.6}
}
\date{}
\maketitle


{\raggedright
    \vspace{1em}
}

\begin{flushleft}
  I. A. M. DATTA \\
  Zap Energy \\
  Seattle, USA \\
  Email: idatta@zap.energy
\end{flushleft}

\begin{flushleft}
  E. T. MEIER \\
  Zap Energy \\
  Seattle, USA \\
\end{flushleft}

\begin{flushleft}
  U. SHUMLAK \\
  Zap Energy / University of Washington \\
  Seattle, USA \\
\end{flushleft}

\section*{\bfseries\MakeTextUppercase{a}\MakeTextLowercase{bstract}}
{\fontsize{9pt}{12pt}\selectfont\hspace{1cm}
  The FuZE sheared-flow-stabilized Z pinch at Zap Energy is simulated using whole-device modeling employing an axisymmetric resistive magnetohydrodynamic formulation implemented within the discontinuous Galerkin WARPXM framework.  Simulations show formation of Z pinches with densities of approximately 10\textsuperscript{22} m\textsuperscript{-3} and total DD fusion neutron rate of 10\textsuperscript{7} per $\mu$s for approximately 2 $\mu$s.  Simulation-derived synthetic diagnostics show peak currents and voltages within 10\% and total yield within approximately 30\% of experiment for similar plasma mass.  The simulations provide insight into the plasma dynamics in the experiment and enable a predictive capability for exploring design changes on devices built at Zap Energy.}

\renewcommand{\footnoterule}{\hrule width \linewidth}

\section{Introduction}\label{sec:introduction}
Whole device modeling of the FuZE shear-flow-stabilized (SFS) Z pinch \cite{Shumlak2020, Zhang2019} at Zap Energy is illuminating key physics of the device and guiding progress towards a breakeven fusion concept with a high-gain power-producing power plant.  Two-dimensional axisymmetric resistive magnetohydrodynamic (MHD) simulations using the discontinuous Galerkin finite element WARPXM framework \cite{Shumlak2011} are compared to experimental measurements from the FuZE device using a suite of synthetic diagnostics.  The modeling unveils the plasma dynamics inside FuZE and enables in-depth investigation of various areas including plasma compression adiabaticity and electrode erosion.  Validated modeling provides a predictive tool for next-step SFS Z pinch experimental design.  \vspace{1em}

\noindent
This paper summarizes simulation results using synthetic diagnostics to compare with experimental analogues.  Section \ref{sec:model} describes the MHD model used while Sec.~\ref{sec:numerical-method} briefly describes the numerical method.  Section \ref{sec:problem-setup} describes the problem setup, Sec.~\ref{sec:results} discusses results, and Sec.~\ref{sec:conclusion} gives conclusions.


\section{Model}\label{sec:model}
For FuZE plasmas, expected temperatures are on the order of 1 keV for a pinch size of about 1 cm with enclosed currents of up to 500 kA \cite{Shumlak2020, Levitt2023}.  In this regime, the ion Larmor radius and ion inertial length are on the millimeter scale.  For the MHD model to be valid, these length scales should be small compared to the system \cite{Meier2021, Toth2017}.  For a 1 keV pinch with 500 kA current, a radius of 1 cm would be on the order of 10 ion Larmor radii, so we proceed with a resistive MHD model given by 
\begin{equation}
  \label{eq:mhd_continuity}
  \frac{\partial \rho}{\partial t} + \nabla\cdot\left(\rho \mathbf{u}\right)
    =
    0
    \mathrm{,}
  \end{equation}
\begin{equation}
  \label{eq:mhd_momentum}
  \frac{\partial  \left(\rho\mathbf{u}\right)}{\partial t} + \nabla\cdot\left(\rho \mathbf{u} \mathbf{u} + p\overline{\overline{I}} + \overline{\overline{\Pi}}\right)
    =
    \left(\omega_{c}\tau\right)\mathbf{J}\times\mathbf{B}
    \mathrm{,}
  \end{equation}
\begin{equation}
  \label{eq:mhd_energy}
  \frac{\partial e_{t}}{\partial t} + \nabla\cdot\left[\left(e + p\right)\mathbf{u} + \overline{\overline{\Pi}}\cdot\mathbf{u} + \mathbf{h}\right]
    =
    -\nabla\cdot\left(\mathbf{E}\times\mathbf{B}\right) - C_{bb}\frac{\rho}{A_{\mathrm{i}}}T^{4}
    \mathrm{,}
\end{equation}  
\begin{equation}
  \label{eq:mhd_faraday}
  \frac{\partial \mathbf{B}}{\partial t} + \nabla\times\mathbf{E}
  =
  0
    \mathrm{,}
\end{equation}  
\begin{equation}
  \label{eq:mhd_total_energy_definition}
  e_{t}
  =
  e + \frac{\mathbf{B}\cdot\mathbf{B}}{2}
  =
  \frac{p}{\gamma - 1} + \frac{\rho\mathbf{u}\mathbf{u}}{2} + \frac{\mathbf{B}\cdot\mathbf{B}}{2}
    \mathrm{,}
\end{equation}  
\begin{equation}
  \label{eq:mhd_current_definition}
  \mathbf{J}
  =
  \frac{\nabla\times\mathbf{B}}{\left(\omega_{c}\tau\right)}
    \mathrm{,}
\end{equation}  
\begin{equation}
  \label{eq:mhd_ohms_law}
  \mathbf{E}
  =
  -\mathbf{u}\times\mathbf{B} + \frac{\left(\nu_{p}\tau\right)}{\left(\omega_{c}\tau\right)}\eta\mathbf{J}
  \equiv
  \mathbf{E}_{\mathrm{Ideal}} +   \mathbf{E}_{\mathrm{Resistive}}
    \mathrm{,}
\end{equation}  
\begin{equation}
  \label{eq:mhd_Pi}
  \overline{\overline{\Pi}}
  =
  -\mu
  \left[\nabla\mathbf{u} + \left(\nabla\mathbf{u}\right)^{T} - \frac{2}{3}\left(\nabla\cdot\mathbf{u}\right)\overline{\overline{I}}\right]
    \mathrm{,}
  \end{equation}
\begin{equation}
  \label{eq:mhd_h}
  \mathbf{h}
  =
  -k\nabla T
  \mathrm{.}
\end{equation}

\noindent
Equations \eqref{eq:mhd_continuity}-\eqref{eq:mhd_h} are written in non-dimensional form with normalized cyclotron frequency $\left(\omega_{c}\tau\right)$ and collision frequency $\left(\nu_{p}\tau\right)$ defined as described in Ref.~\cite{Datta2023}.  The equations are solved, as described in the next section, in two-dimensional ($r$-$z$ plane) axisymmetric form where $\frac{\partial}{\partial\theta}=0$.\vspace{1em}

\noindent
A blackbody radiation model \cite{Mikellides2011} is added in Eq.~\eqref{eq:mhd_energy}.  Viscosity and thermal conduction terms are added as well, as seen in Eqs.~\eqref{eq:mhd_momentum}, \eqref{eq:mhd_energy}, \eqref{eq:mhd_Pi}, and \eqref{eq:mhd_h}.  These terms are required to match the simulation dynamics with experiment, as described further in Sec.~\ref{sec:results}.




\section{Numerical Method}\label{sec:numerical-method}
The MHD model described in Eqs.~\eqref{eq:mhd_continuity}-\eqref{eq:mhd_h} can be combined and expressed compactly as
\begin{equation}
  \label{eq:conservative_form}
  \frac{\partial \mathbf{q}}{\partial t}
  +
  \nabla\cdot\overline{\overline{F}}
  =
  \mathbf{S}
  \mathrm{.}
\end{equation}

\noindent
The equations are solved using the discontinuous Galerkin method within the WARPXM framework developed at the University of Washington and Zap Energy \cite{Shumlak2011}.  An overview of the method and implementation is found in Refs.~\cite{Shumlak2011}, \cite{Meier2021}, and \cite{Datta2023}.  The discontinuous Galerkin method is well suited for solving the predominantly hyperbolic MHD equations on unstructured meshes using simplex finite elements, allowing for simulation of the FuZE geometry.  In this work, a cylindrical formulation of the MHD equations described in Sec.~\ref{sec:model} is used, the  numerical handling of which is described in Ref.~\cite{Meier2021}.  Artificial dissipation is also added to stabilize solutions \cite{VonNeumann1950}.\vspace{1em}

\noindent
A circuit model is implemented to capture the interaction between the driving current bank and the plasma in FuZE in a self-consistent manner.  The circuit is modeled by including the plasma dynamics as a time-dependent impedance and associated load voltage as a circuit element.  The current is driven by the circuit with prescribed capacitance, inductance, resistance, charge voltage, and trigger time. 



\section{Problem Setup}\label{sec:problem-setup}
\begin{figure}[htbp]
\centering
\includegraphics[width=1.0\textwidth]{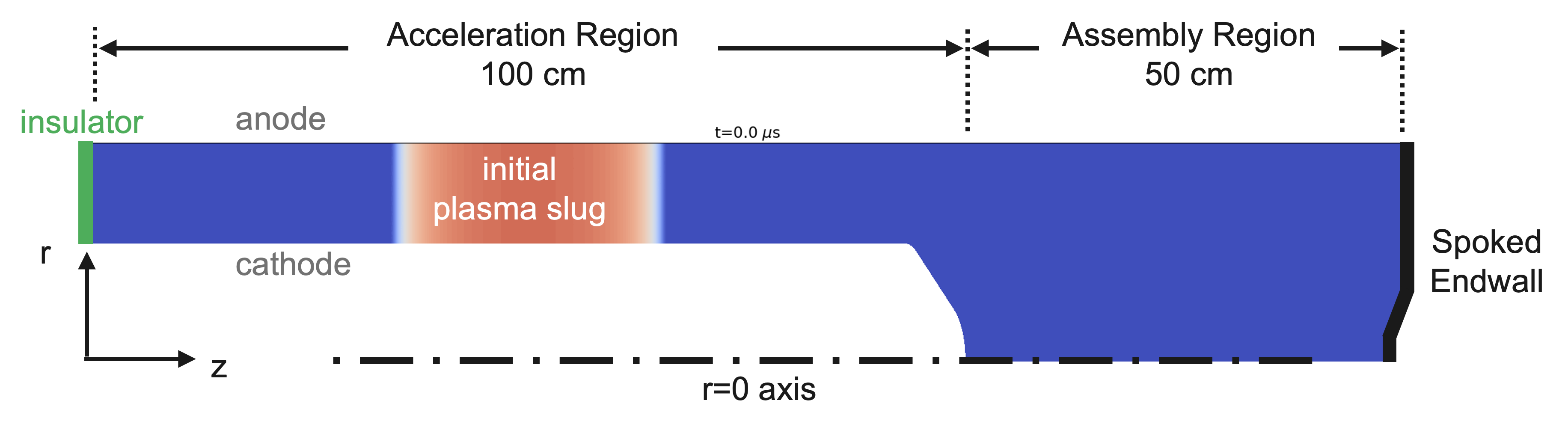}
\captionsetup{font=it}
\caption{Two-dimensional $r$-$z$ plane slice of FuZE used as the simulation domain.  Gas is puffed into the coaxial acceleration region where it is ionized by an electrical power supply between the cathode (inner electrode) and anode (outer electrode) and pushed into the assembly region to form a pinch.  The simulation assumes the puffed gas is fully ionized as an initial plasma slug at $t=0$.  Note that the radial dimension is stretched roughly by 3-to-1.}
\label{fig:machine_drawing}
\end{figure}

The simulations performed aim to match conditions for a set of experimental operating parameters on FuZE as shown in Fig.~\ref{fig:machine_drawing}.  Triangular finite elements are used with a quadratic solution basis.  A slug of plasma of 0.5 mg in the acceleration region is initialized where the valves inject gas.  A volumetric injection of plasma is also input at a rate of 3 mg / ms at the valve location, mimicking breakdown of residual neutral gas and ongoing valve injection. \vspace{1em}

\noindent
Simulated initial mass is chosen by following the trajectory of the magnetic field data for an experimental plasma pulse (\#230518072) and calculating the corresponding plasma mass.  Valve mass injection rates are also estimated, as described in Sec.~\ref{sec:mass-determ-from}.  A series RLC circuit is used to simulate the driving capacitor bank for FuZE, with circuit values $C=222 ~\mu \mathrm{F}$, $L=286 \mathrm{~nH}$, and $R = 1.5 \mathrm{~m}\Omega$.  At $t=0$, the capacitor charged to 25 kV is discharged.  The plasma load is calculated as a voltage drop measured along the insulator of the device between the cathode (inner electrode) and anode (outer electrode).  The gap voltage is measured upstream of the insulator ahead of a 70 nH region of inductance (see Fig.~1 in Ref.~\cite{Shumlak2023}), so to compare with experiment, it is calculated as
$
  V_{\mathrm{gap}}
  =
  -\int\limits_{\text{cathode}}^{\text{anode}}E_{r}dr  
  + \left(\mathrm{70~nH}\right)\dot{I}
  \mathrm{.}
  $
  \vspace{1em}
  
\noindent  
The circuit current, converted to an azimuthal magnetic field, imposes a boundary condition on the plasma at the insulator, where the axial velocity is allowed to float if into the domain and set to zero otherwise.  At $r=0$, axisymmetric boundary conditions are employed, as described in Ref.~\cite{Meier2021}.  All other boundaries are treated as free-slip conducting walls that are thermally insulating by imposing zero-normal density and pressure gradients.

\subsection{Mass determination from slug and valve injection models}\label{sec:mass-determ-from}
The mass of the plasma in experiment is estimated using a combination of magnetic probes along the outer electrode and gas valve injection formulas.  Using the magnetic probes, the motion of an assumed slug of plasma injected by the gas puff valves is tracked to estimate its mass.  Using this slug model, the plasma mass is determined using the force \cite{Stepanov2020, Jahn2006},
\begin{equation}
  \label{eq:snowplow_force}
  F(t)
  =
  \frac{L^{'}I(t)^{2}}{2}
  =
  m
  \ddot{z}
  \mathrm{.}
\end{equation}
Dividing Eq.~\eqref{eq:snowplow_force} by $m$ and integrating twice to time, $t_{d}$, leads to an equation for distance traveled by the plasma,
\begin{equation}
  \label{eq:snowplow_distance}
  d
  =
  \frac{L^{'}}{2m}\iint\limits_{0}^{t_{d}}I^{2}dt
  \mathrm{.}
\end{equation}

\begin{wrapfigure}[16]{L}{0.50\textwidth}
    \centering
    \includegraphics[width=0.50\textwidth]{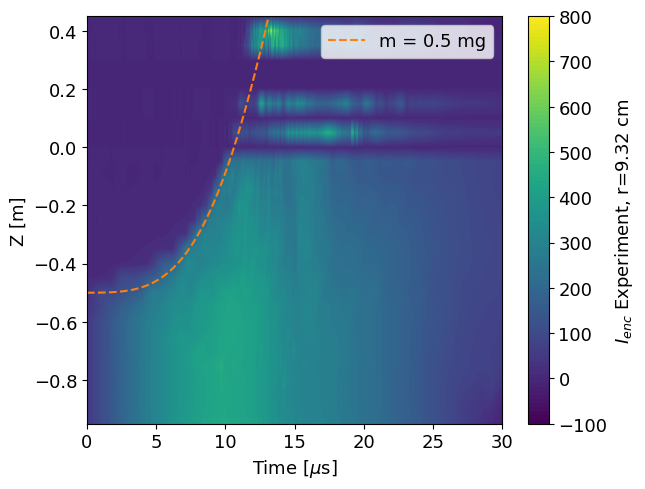}
\captionsetup{font=it}
\caption{Enclosed current in experiment as measured by magnetic probes along the outer electrode as a function of time.  A few probes in the assembly region did not register data for this pulse.}
\label{fig:230518072_enclosed_current}
\end{wrapfigure}

\noindent
In the experiment, the times taken for the plasma slug to move from the gas puff valves to the magnetic probes along the outer electrode are known by measuring when each probe records a jump in magnetic field.  A space-time plot of this magnetic probe data as shown in Fig.~\ref{fig:230518072_enclosed_current} indicates where and when along the outer electrode the jumps occur.  With $d$ and $t_{d}$ known for various magnetic probe locations, $m$ is calculated using Eq.~\eqref{eq:snowplow_distance}.  The dashed orange line in Fig.~\ref{fig:230518072_enclosed_current} corresponds to $m = 0.5$ mg for the particular plasma pulse.\vspace{1em}

\noindent
A combination of gas valve injection and a deflagration process ionizing remaining neutral gas sustains plasma flow to the pinch \cite{Stepanov2020}.  To account for that effect, a mass injection model is needed.  A simple model assumes gas flows out of the valve at the speed of sound such that
\begin{equation}
  \label{eq:valve_diff_eq}
  \frac{dm}{dt}
  =
  V\frac{d\rho}{dt}
  =
  -\rho c_{s}A
  =
  -A \rho\sqrt{\frac{\gamma p }{\rho}}
  \mathrm{.}
\end{equation}
Normalizing by non-dimensional parameters, $\tilde{\rho}=\rho/\rho_{0}$ and $\tilde{t}=t/\tau_{v}$ where $\tau_{v}=V/c_{s0}A$ is the timescale of the venting valve and assuming adiabatic expansion such that $\tilde{p}=p/p_{0}=\left(\rho/\rho_{0}\right)^{\gamma}={\tilde{\rho}}^{\gamma}$ leads to $\frac{d\tilde{\rho}}{d\tilde{t}} = -\tilde{\rho}^{\frac{\gamma + 1}{2}}$.
The solution for $\tilde{\rho}\left(0\right)=1$ is $\tilde{\rho} = \left(\frac{\gamma - 1}{2}\tilde{t} + 1\right)^{\frac{2}{1-\gamma}}$,
which describes the normalized density in the valve.  The normalized density entering the device is thus $\tilde{\rho} = 1 - \left(\frac{\gamma - 1}{2}\tilde{t} + 1\right)^{\frac{2}{1-\gamma}}$.
Substitution of the dimensional values leads to 
an expression in terms of initial density, $\rho_{0}$, which is recast in terms of valve pressure, $p$, using the ideal gas law for deuterium (assuming $V=1 \mathrm{~cm^{3}}$ and $T=300\mathrm{~K}$).  A reference mass calculated for a pressure of 100 PSIA is $M_{0,\mathrm{ref}}=\rho_{0,\mathrm{ref}}V = \ 1.11 \mathrm{~mg}$.  For different pressures, this reference mass is scaled linearly, leading to mass as a function valve pressure and opening time as 
\begin{equation}
  \label{eq:valve_diff_eq_normalized_solution_device_dimensional_vs_pressure}
  m
  =
  N_{v}M_{0,\mathrm{ref}}\frac{p}{100}
  \left[
    1 -
    \left\{
      1 + \frac{\gamma - 1}{2}\left(\frac{t - \Delta}{\tau_{v}}\right)
    \right\}^{\frac{2}{1 - \gamma}}
  \right]
  \mathrm{,}
\end{equation}
where $\Delta$ is a timing delay between the valve trigger and opening, $N_{v}$ is the number of valves used, and $p$ is measured in PSIA.  For the experimental plasma pulse in Fig.~\ref{fig:230518072_enclosed_current}, $N_{v} = 4$, $p=149.1$ PSIA, and $t=0.26$ ms.  For deuterium ($\gamma=7/5$, $T=300\mathrm{~K}$), the sound speed is 934 m/s leading to $\tau_{v}=1.070$ ms (assuming $A = 1 \mathrm{~mm^{3}}$).  Equation \eqref{eq:valve_diff_eq_normalized_solution_device_dimensional_vs_pressure} is solved for $\Delta$ using $m=0.5$ mg, yielding $0.175$ ms.  Equation \eqref{eq:valve_diff_eq_normalized_solution_device_dimensional_vs_pressure} is then differentiated with respect to time, yielding $\dot{m}=3$ mg / ms.  Such an injection rate may be considered a low estimate, since deflagration is expected to make a significant contribution as an upstream plasma source.  Modeling with neutrals to capture deflagration is the topic of ongoing research.\vspace{1em}

\noindent
A simulation (\#s230523033) is thus run to 30 $\mu$s matching experimental pulse settings with 0.5 mg slug mass and 3 mg / ms injection rate.  The initial slug of plasma sits in the acceleration region where the valves are located, as shown in Fig.~\ref{fig:machine_drawing}.

\section{Results}\label{sec:results}
\begin{figure}[htbp]
    \centering
    \begin{minipage}[t]{0.49\textwidth}
        \centering
        \includegraphics[width=1.00\textwidth]{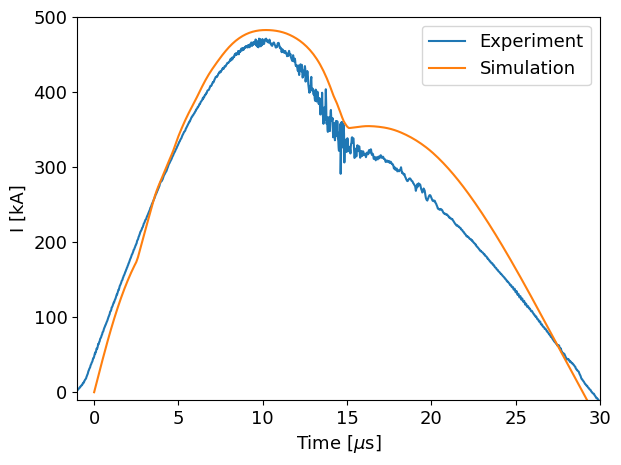} 
        \captionsetup{font=it}
        \caption{Current comparison between experiment (blue) and simulation (orange).  Good agreement is found especially in earlier times.}
        \label{fig:I_s230523033_and_230518072}        
    \end{minipage}\hfill
    \begin{minipage}[t]{0.49\textwidth}
        \centering
        \includegraphics[width=1.00\textwidth]{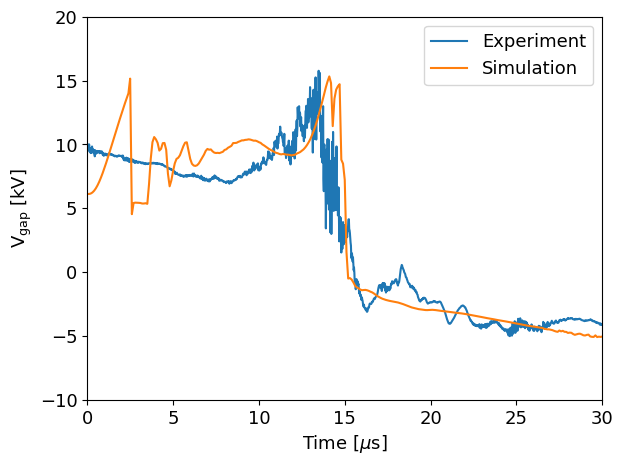} 
        \captionsetup{font=it}
        \caption{$V_{\mathrm{gap}}$ comparison between experiment (blue) and simulation (orange).  Similar values and temporal profiles are seen, with jumps early in time in simulation being an artifact of the fully-ionized MHD assumptions.}
        \label{fig:v_gap_s230523033_and_230518072}        
    \end{minipage}
  \end{figure}

Traces of current and voltage are shown in Figs.~\ref{fig:I_s230523033_and_230518072} and \ref{fig:v_gap_s230523033_and_230518072} with comparisons to experimental discharge \#230518072.  In all experimental plots, timings are shifted by 4.2 $\mu$s to when the current rise starts.\vspace{1em}

\noindent
Figure \ref{fig:I_s230523033_and_230518072} compares experimental capacitor bank current output with that of the simulation coupling the equivalent RLC series circuit described in Sec.~\ref{sec:problem-setup} with the MHD plasma.  The currents reach similar peaks before dropping, with slightly lower currents later in the experimental pulse compared to simulation. \vspace{1em}

\noindent  
Figure \ref{fig:v_gap_s230523033_and_230518072} shows a comparison of $V_{\mathrm{gap}}$ between experiment and simulation.  Early in time, the simulation exhibits jumps in voltage not present in experiment, an artifact of the fully ionized MHD assumptions of the initial slug.  The incoming plasma flux is reflected by the slug, causing the jumps in $V_{\mathrm{gap}}$.  Nonetheless, the gap voltage values in experiment and simulation are similar and have comparable temporal profiles with the drop in voltage at around 15 $\mu$s.  The voltage drop is associated with pileup of plasma and flux at the endwall that drives a wave of magnetic flux back into the acceleration region where it interacts with the circuit.\vspace{1em}

\begin{figure}[htbp]
     \centering
         \includegraphics[width=1.0\textwidth]{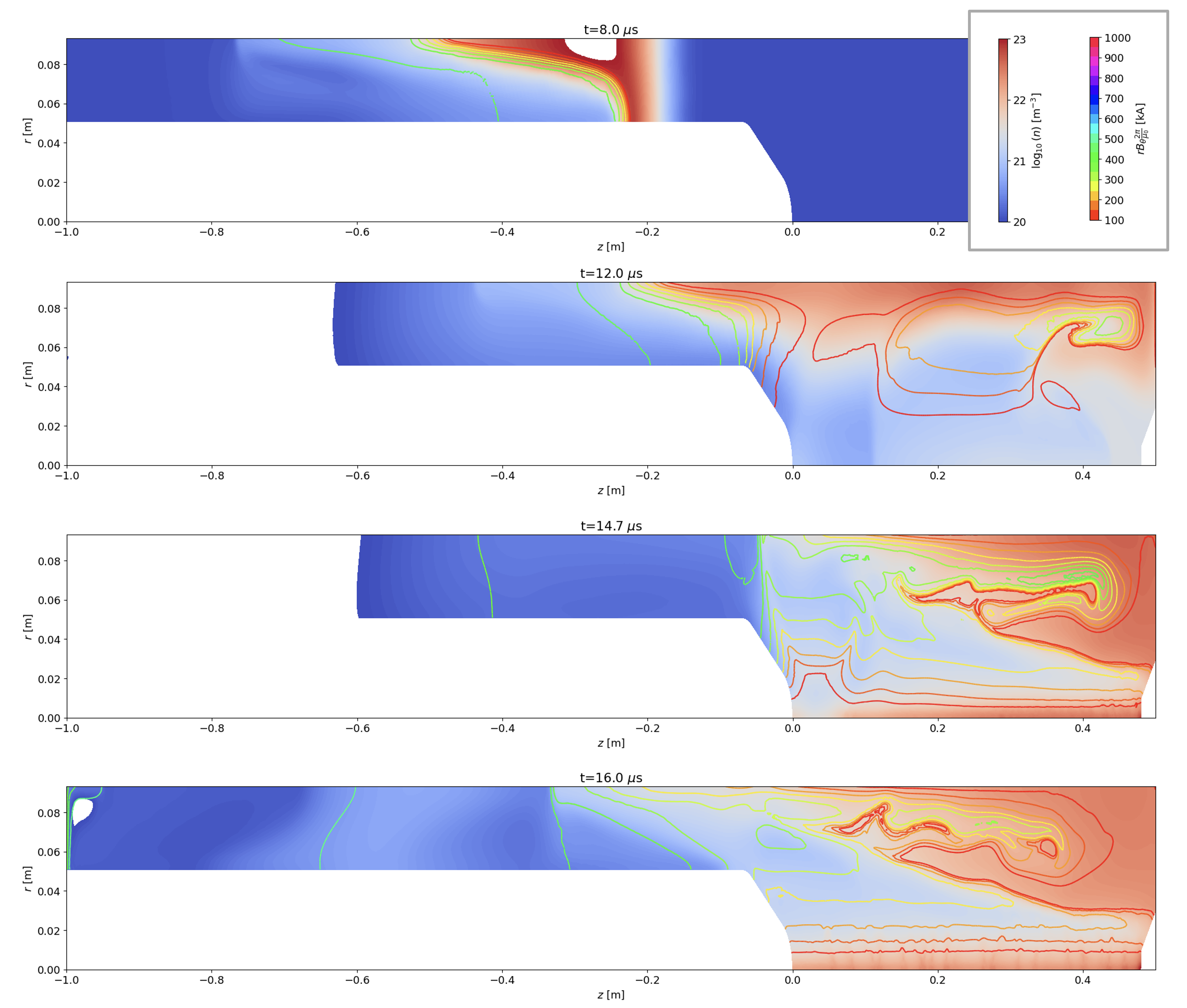}
       \captionsetup{font=it}       
        \caption{Plasma density at 8 $\mu$s, 12 $\mu$s, 14.7 $\mu$s (maximum total domain-integrated yield rate time), and 16 $\mu$s, overlaid with lines of enclosed current.  The plots show current sheet canting and blowby as the magnetic flux pushes the plasma slug into the assembly region.  Plasma thermal energy and magnetic flux accumulate at the endwall, causing a rebound behavior that completes pinch formation.  The 16-$\mu$s plot shows the pinch formed on axis with noticeable $m=0$ activity associated with reduced sheared flow.}
        \label{fig:s230523033_n}
      \end{figure}

\noindent
Figure \ref{fig:s230523033_n} shows the evolution of plasma density along with lines of enclosed current.  The plasma slug is pushed from the acceleration region to the assembly region through the $\mathbf{J}\times\mathbf{B}$ force imposed by the circuit current.  A true slug motion would be indicated by the lines of radial current through the plasma as it moves through the acceleration region.  At $t=8$ $\mu$s, deviation from the 1D slug model is apparent, as axial components of the radial contours are seen near the outer electrode.  Such behavior represents a blowby phenomenon where the imposed current does not fully couple to the plasma \cite{Cassibry2006}.  The viscous ($\nu = \mu/\rho$) and heat ($\kappa = k A_{\mathrm{i}} / \rho$) diffusivities in this simulation are set to 200 m\textsuperscript{2} / s which lead to the good agreement with experimental values shown in Figs.~\ref{fig:I_s230523033_and_230518072} and \ref{fig:v_gap_s230523033_and_230518072}.  Simulations with increased viscous and heat diffusivities show reduced blowby and increased peak current and voltage.  After the slug reaches the assembly region and the flux rebounds after hitting the endwall, a pinch then forms on axis, as seen in the high density on-axis regions in the 14.7 $\mu$s and 16 $\mu$s plots.  \vspace{1em}

\begin{figure}[htbp]
     \centering
     \includegraphics[width=1.0\textwidth]{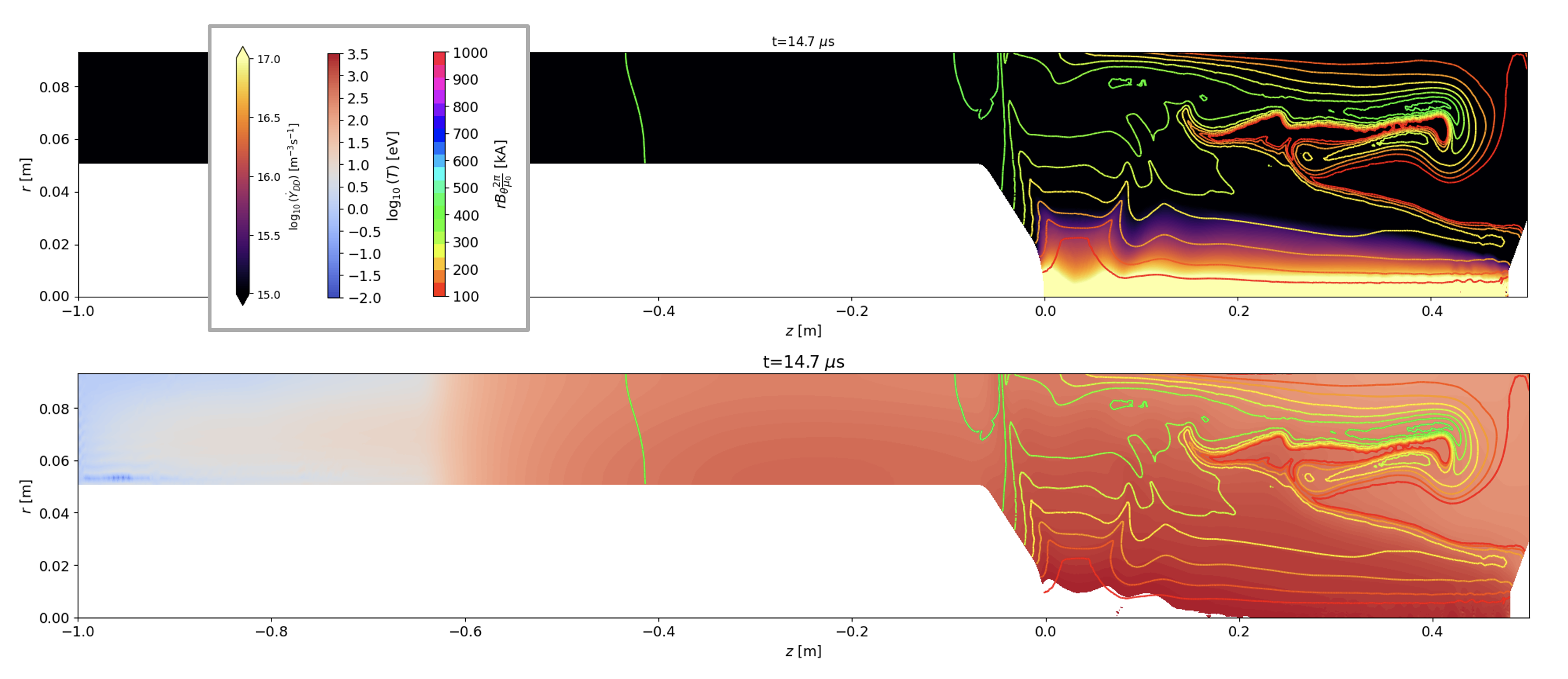}
       \captionsetup{font=it}       
        \caption{Neutron yield rate calculated from Bosch-Hale formulas for thermonuclear fusion \cite{Bosch1992} (top) and plasma temperature (bottom) at 14.7 $\mu$s, the time of maximum total domain-integrated yield rate.  Lines of enclosed current are also shown.  The plasma reaches above 3 keV near the nosecone on axis in the assembly region.  This high temperature combined with collimated density on axis lead to the region of thermonuclear fusion seen in the neutron yield rate plot.}
        \label{fig:s230523033_bangtime}
\end{figure}

\noindent 
Figure \ref{fig:s230523033_bangtime} shows the volumetric neutron yield rate calculated using Bosch-Hale thermonuclear fusion formulas \cite{Bosch1992} at the time of maximum total yield as well as plasma temperature.  The thermonuclear yield is a function of the density and temperature.  The combination of high temperature on axis above 3 keV combined with on-axis densities above 10\textsuperscript{22} m\textsuperscript{-3} leads to the on-axis yield rate, indicating the existence of a Z pinch in the simulation. \vspace{1em}
      
\begin{figure}[htbp]  
    \centering
    \includegraphics[width=0.475\textwidth]{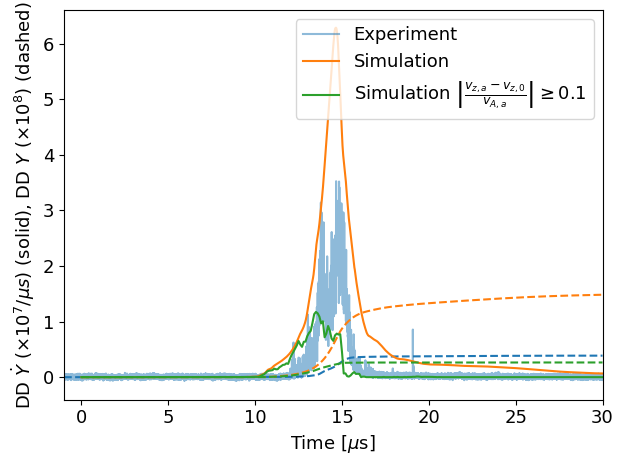}
\captionsetup{font=it}
\caption{Total yield comparison between experiment and simulation.  The experimental yield rate (solid blue) is found using plastic scintillators calibrated such that the time-integrated yield agrees with Lanthanum Bromide detectors.  Total volume yield rate from simulation is calculated by integrating the volumetric yield as shown in Fig.~\ref{fig:s230523033_bangtime} (solid orange).  The solid green line also shows the simulation volume yield rate after removing yield in regions a pinch would be assumed unstable.  Dashed lines integrate the yield rate to determine the total yield.}
\label{fig:yield_s230523033_and_230518072}
\end{figure}

\noindent
Figure \ref{fig:yield_s230523033_and_230518072} shows the total integrated yield rate in the simulation versus experimental data.  The experimental data is of fast plastic scintillator measurements scaled to instantaneous neutron yield rates using total yield measurements by Lanthanum Bromide detectors.  The simulation data is an analogous synthetic diagnostic calculating the full-domain integral of the simulated volumetric yield rate (the top plot in Fig.~\ref{fig:s230523033_bangtime}).  The dashed lines are integrations of the total yield rates over time.  The difference in simulation and experimental yield indicates an overprediction in simulation by about 300\%.  An adjustment to the simulation yield rate calculation is made by removing yield where the difference between velocity at the pinch edge, the radius $a$ where $p(a) = p_{\mathrm{max}}/4$ \cite{Meier2021}, and the axis is below 10\% of the Alfv\'{e}n speed at $a$.  The correction removes yield for low sheared flows assumed to be $m=1$ unstable \cite{Shumlak1995, Meier2021}.  The total yield calculated from the correction is about 70\% of the experimental yield.\vspace{1em}

\noindent
In simulations without viscous and heat diffusivities (not shown), the yield rate increases early compared to the experiment.  Increased blowby in such cases lead to earlier flux rebound from the endwall and thus earlier neutron production.  This again suggests some viscous dissipation is needed to match experimental results.  Additionally, simulations with no radiation have higher neutron production at times after the peak, unlike in the experiment where neutron production drops significantly after about 16 $\mu$s.  The addition of the blackbody radiation term reduces this late time yield in the simulation, similar to experiment.  The physical value for the blackbody radiation constant used in simulations is $C_{bb}= 4.559 \times 10^{52} \mathrm{~J}^{-3}\mathrm{s}^{-1}$.\vspace{1em}

\noindent
Further work to simulate the plasma dynamics with more realistic viscosity and radiation models are ongoing.  This includes the consideration of Braginskii \cite{Braginskii1965} terms as well as bremsstrahlung and line-radiation models \cite{Rognlien2002}.  Capturing the true nature of the endwall flux rebound is also being considered.  Current modeling does not include the possibility of outflow through the endwall which has a spoked transparent geometry \cite{Claveau2020} or dissipation due to ion recycling and re-ioniziation \cite{Stangeby2000} that may weaken the flux rebound.  However, it is unlikely that there is full outflow \cite{Claveau2020} so, in future work, boundary conditions and related physics (including 3D effects) will be developed to capture realistic endwall behavior. \vspace{1em}

\begin{figure}[htbp]
    \centering
    \begin{minipage}[t]{0.49\textwidth}
        \centering
        \includegraphics[width=1.00\textwidth]{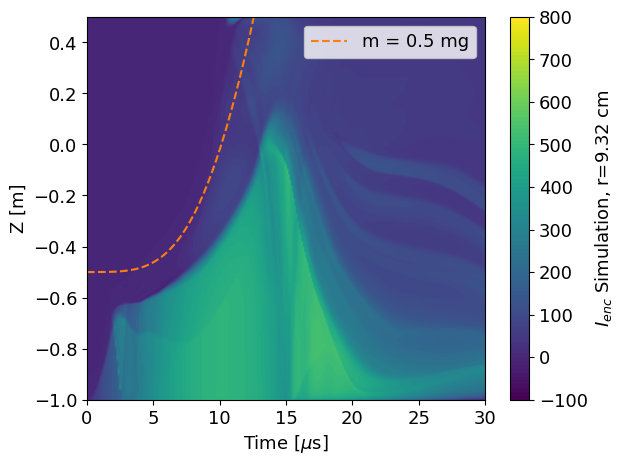} 
        \captionsetup{font=it}
        \caption{Enclosed current in simulation at outer electrode radius.  Large eddy currents in simulation reduce these values in the assembly region.}
        \label{fig:Ienc_sim_s230523033_0p0932}        
    \end{minipage}\hfill
    \begin{minipage}[t]{0.49\textwidth}
        \centering
        \includegraphics[width=1.00\textwidth]{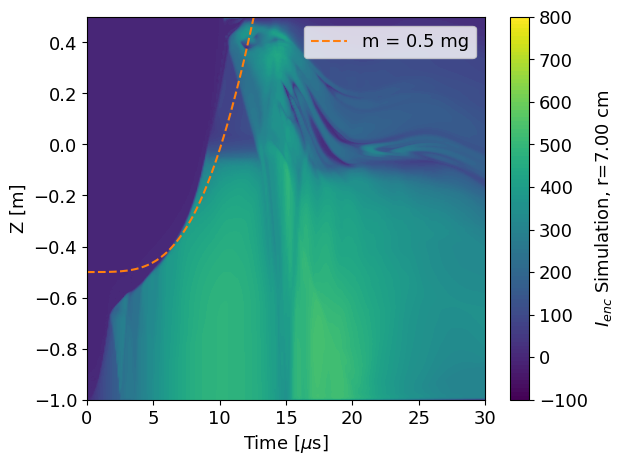} 
        \captionsetup{font=it}
        \caption{Enclosed current in simulation at $r = 7$ cm.  At this radius, some of the eddy currents are removed.  The 0.5 mg trajectory line matches better than in Fig.~\ref{fig:Ienc_sim_s230523033_0p0932}.}
        \label{fig:Ienc_sim_s230523033_0p07}        
    \end{minipage}
  \end{figure}

\noindent
Figures \ref{fig:Ienc_sim_s230523033_0p0932} and \ref{fig:Ienc_sim_s230523033_0p07} show simulation synthetic analogues of the space-time enclosed current plot shown in Fig.~\ref{fig:230518072_enclosed_current}.  Figure \ref{fig:Ienc_sim_s230523033_0p0932} shows the enclosed current at the outer electrode with the 0.5 mg slug trajectory superimposed.  The lack of agreement with Fig.~\ref{fig:230518072_enclosed_current} is understood by observing Fig.~\ref{fig:s230523033_n}.  Between 12 $\mu$s and 16 $\mu$s, a clear detachment of enclosed current is seen along the outer electrode due to the formation of eddy currents, which reduce the total enclosed current measured.  In contrast, the experiment shows strong magnetic fields at the outer electrode in the assembly region as seen in Fig.~\ref{fig:230518072_enclosed_current}.  For the simulation, a similar plot is constructed at a smaller radius inside of the eddy currents.  Choosing 7 cm as shown in Fig.~\ref{fig:Ienc_sim_s230523033_0p07} indicates a magnetic space-time structure more in line with the experiment in Fig.~\ref{fig:230518072_enclosed_current}, with the 0.5 mg trajectory close to the magnetic field wave front.

\vspace{1em}




\section{Conclusion}\label{sec:conclusion}
Comparisons of WARPXM simulations using two-dimensional axisymmetric resistive MHD with the FuZE experiment at Zap Energy show a remarkable level of agreement.  Comparisons using a suite of synthetic diagnostics illuminate these agreements while suggesting areas that modeling does not fully capture.  However, these results show that the 2D axisymmetric MHD model can be used as a predictive tool for exploring alternative geometries, electrode configurations, driving circuit parameters, and gas fill profiles which can accelerate progress toward breakeven fusion gain.  Improvements to modeling are possible and will improve prediction accuracy.  Extension of the MHD model to 3D can capture non-axisymmetric effects leading to $m=1$ instability.  The inclusion of multi-fluid physics \cite{Shumlak2003, Shumlak2011, Meier2021} can help to accurately capture blowby, sheath effects, and instability growth rates.  Multi-species continuum kinetic modeling can allow for further accuracy of sheath modeling and pinch dynamics at small pinch radius where the magnetic field goes to zero \cite{Reddell2016}, with hybrid approaches helping to speed up such simulations \cite{Ho2018, Datta2023}.  Other advancements include the addition of a plasma-neutral model \cite{Meier2012} to capture plasma deflagration \cite{Stepanov2020} and more accurate boundary conditions that can capture heat and particle fluxes to walls, and corresponding erosion and impurity generation \cite{Rognlien2002}.  These advancements are actively being developed to help capture the plasma dynamics of FuZE and other devices being built at Zap Energy \cite{Levitt2023}.

\section*{\hfill \textbf{ACKNOWLEDGEMENTS} \hfill}
The information, data, or work presented herein was funded in part by the Advanced Research Projects Agency – Energy (ARPA-E), U.S. Department of Energy, under Award Nos. DE-AR-0000571, DE-AR-0001010, DE-AR-0001260 and by the Air Force Office of Scientific Research under Grant No. FA9550-15-1-0271.  This research used resources of the National Energy Research Scientific Computing Center (NERSC), a U.S. Department of Energy Office of Science User Facility located at Lawrence Berkeley National Laboratory, operated under Contract No. DE-AC02-05CH11231.

\section*{\hfill \textbf{REFERENCES}\hfill}
\bibliographystyle{phy-bstyles/iaea_iman}
\patchcmd{\thebibliography}{\section*{\refname}}{}{}{}
\bibliography{References}

\end{document}